\documentstyle[epsfig,multicol,pre,aps]{revtex}
\begin{document}
\draft

\title{Bicritical Behavior of Period Doublings in
       Unidirectionally-Coupled Maps}

\author{Sang-Yoon Kim
\footnote{Electronic address: sykim@cc.kangwon.ac.kr}
        }
\address{
 Department of Physics\\ Kangwon National University\\
 Chunchon, Kangwon-Do 200-701, Korea
 }

\maketitle

\begin{abstract}
We study the scaling behavior of period doublings in two
unidirectionally-coupled one-dimensional maps near a bicritical point
where two critical lines of period-doubling transition to chaos in
both subsystems meet. Note that the bicritical point corresponds
to a border of chaos in both subsystems. For this bicritical case,
the second response subsystem exhibits a new type of non-Feigenbaum
critical behavior, while the first drive subsystem is in the
Feigenbaum critical state. Using two different methods, we make the
renormalization group analysis of the bicritical behavior and
find the corresponding fixed point of the renormalization
transformation with two relevant eigenvalues. The scaling factors
obtained by the renormalization group analysis agree well with those
obtained by a direct numerical method.
\end{abstract}


%
%
\begin{multicols}{2}
\section{Introduction}
\label{sec:Int}

Period-doubling transition to chaos has been extensively studied
in a one-parameter family of one-dimensional (1D) unimodal maps,
\begin{equation}
  x_{t+1} = 1 - A x_t^2,
\label{eq:1DM}
\end{equation}
where $x_t$ is a state variable at a discrete time $t$.
As the control parameter $A$ is increased, the 1D map undergoes
an infinite sequence of period-doubling bifurcations accumulating
at a critical point $A_c$, beyond which chaos sets in. Using a
renormalization group (RG) method, Feigenbaum \cite{Feigenbaum} has
discovered universal scaling behavior near the critical point $A_c$.

Here we are interested in the period doublings in a system
consisting of two 1D maps with a one-way coupling,
\begin{equation}
  x_{t+1} = 1- A x_t^2,\;\;
  y_{t+1}= 1 - B y_t^2 -C x_t^2,
\label{eq:UCM}
\end{equation}
where $x$ and $y$ are state variables of the first and second
subsystems, $A$ and $B$ are control parameters of the subsystems, and
$C$ is a coupling parameter. Note that the first (drive) subsystem
acts on the second (response) subsystem, while the second subsystem
does not influence the first subsystem. This kind of
unidirectionally-coupled 1D maps are used as a model for open flow
\cite{OF}. A new kind of non-Feigenbaum scaling behavior was found in
a numerical and empirical way near a bicritical point $(A_c,B_c)$
where two critical lines of period-doubling transition to chaos in
both subsystems meet \cite{Kuz1}. For this bicritical case, a RG
analysis was also developed and the corresponding fixed point,
governing the bicritical behavior, was numerically obtained by
directly solving the RG fixed-point equation using a polynomial
approximation \cite{Kuz2}. In this paper, using two different
methods, we also make the RG analysis of the bicriticality,
the results of which agree well with those of previous works.

This paper is organized as follows. In Sec.~\ref{sec:SB} we study
the scaling behavior near a bicritical point $(A_c,B_c)$,
corresponding to a border of chaos in both subsystems, by directly
following a period-doubling sequence converging to the point
$(A_c,B_c)$ for a fixed value of $C$. For this bicritical case, a
new type of non-Feigenbaum critical behavior appears in the second
subsystem, while the first subsystem is in the Feigenbaum critical
state. Employing two different methods, we make the RG analysis of
the bicritical behavior in Sec.~\ref{sec:RA}. To solve the RG
fixed-point equation, we first use an approximate truncation
method \cite{TM}, corresponding to the lowest-order polynomial
approximation. Thus we analytically obtain the fixed point,
associated with the bicritical behavior, and its relevant
eigenvalues. Compared with the previous numerical results
\cite{Kuz2}, these analytic results are not bad as the
lowest-order approximation. To improve accuracy, we also employ
the ``eigenvalue-matching'' RG method \cite{EM}, equating the
stability multipliers of the orbit of level $n$ (period $2^n$) to
those of the orbit of the next level $n+1$. Thus we numerically
obtain the bicritical point, the parameter and orbital scaling
factors, and the critical stability multipliers. We note that the
accuracy is improved remarkably with increasing the level $n$.
Finally, a summary is given in Sec.~\ref{sec:Sum}.

\section{Scaling Behavior near The Bicritical Point}
\label{sec:SB}

In this section we fix the value of the coupling parameter by setting
$C=0.45$ and directly follow a period-doubling sequence converging
to the bicritical point $(A_c,B_c)$, which corresponds to a border of
chaos in both subsystems. For this bicritical case, the second
subsystem exhibits a new type of non-Feigenbaum critical behavior,
while the first subsystem is in the Feigenbaum critical state.

The unidirectionally-coupled 1D maps (\ref{eq:UCM}) has many
attractors for fixed values of the parameters \cite{MS}. For the
case $C=0$, it breaks up into the two uncoupled 1D maps. If they both
have stable orbits of period $2^k$, then the composite system has
$2^k$ different stable states distinguished by the phase shift between
the subsystems. This multistability is preserved when the coupling is
introduced, at least while its value is small enough. Here we study
only the attractors whose basins include the origin $(0,0)$. Such
attractors become in-phase when $A=B$ and $C=0$.

Stability of an orbit with period $q$ is determined by its stability
multipliers,
\begin{equation}
\lambda_1 = \prod_{t=1}^{q} -2 A x_t, \;\; \lambda_2 =
\prod_{t=1}^{q} -2 B y_t.
\label{eq:SM}
\end{equation}
Here $\lambda_1$ and $\lambda_2$ determine the stability of the first
and second subsystems, respectively. An orbit becomes stable when the
moduli of both multipliers are less than unity, i.e., $-1 < \lambda_i
< 1$ for $i=1,2$.

\noindent
\begin{minipage}{\columnwidth}
\begin{figure}
\centerline{ 
 \epsfig{file={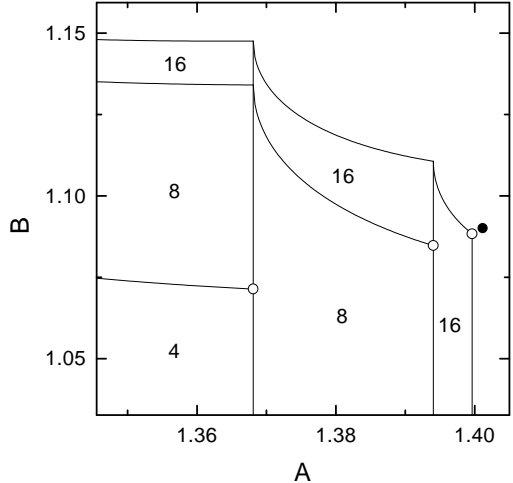}, width=\columnwidth}}
\vspace{-5cm}
\caption{Stability diagram of the periodic orbits born via
 period-doubling bifurcations for $C=0.45$. The numbers in
 the different regions represent the period of motion in the second
 subsystem. The open circle also denotes the point, corresponding
 to a threshold of instability in both subsystems, where
 $\lambda_1 =-1$ and $\lambda_2 =-1$. Such open circles accumulate
 to the bicritical point, denoted by the solid circle, which
 corresponds to a border of chaos in both subsystems.
 For other details, see the text.
     }
\label{fig:SD}
\vspace{0.5cm}
\end{figure}
\end{minipage}

Figure \ref{fig:SD} shows the stability diagram of periodic orbits
for $C=0.45$. As the parameter $A$ is increased, the first
subsystem exhibits a sequence of period-doubling bifurcations at
the vertical straight lines, where $\lambda_1 = -1$. For small
values of the parameter $B$, the period of oscillation in the
second subsystem is the same as that in the first subsystem, as in
the case of forced oscillation. As $B$ is increased for a fixed
value of $A$, a sequence of period-doubling bifurcations occurs in
the second subsystem when crossing the non-vertical lines where
$\lambda_2=-1$. The numbers inside the different regions denote
the period of the oscillation in the second subsystem.

We consider a pair of the parameters $(A_n,B_n)$, at which the
periodic orbit of level $n$ (period $2^n$) has the stability
multipliers $\lambda_{1,n}=\lambda_{2,n}=-1$. Hence, the point
$(A_n,B_n)$ corresponds to a threshold of instability in both
subsystems. Some of such points are denoted by the open circles in
Fig.~1. Then such a sequence of $(A_n,B_n)$ converges to the
bicritical point $(A_c,B_c)$, corresponding to a border of chaos
in both systems, with increasing the level $n$. The bicritical
point is denoted by the solid circle in Fig.~1. To locate the
bicritical point with a satisfactory precision, we numerically
follow the orbits of period $q=2^n$ up to level $n=21$ in a
quadruple precision, and obtain the sequences of both the
parameters $(A_n,B_n)$ and the orbit points $(x_n,y_n)$
approaching the origin. We first note that the sequences of $A_n$
and $x_n$ in the first subsystem are the same as those in the 1D
maps \cite{Feigenbaum}. Hence, only the sequences of $B_n$ and
$y_n$ in the second subsystem are given in Table \ref{tab:SPO}.

\noindent
\begin{minipage}{\columnwidth}
\begin{table}
\caption{Sequences of the parameter and the orbit point, $\{ B_n \}$
         and $\{ y_n \}$, in the second subsystem.
        }
\label{tab:SPO}
\begin{tabular}{ccc}
    $n$ & $B_n$ & $y_n$   \\
\tableline
10 & 1.090\,088\,955\,364 &    5.019\,189\,$\times$$10^{-3}$  \\
11 & 1.090\,092\,109\,910 &   -3.333\,775\,$\times$$10^{-3}$ \\
12 & 1.090\,093\,416\,851 &    2.214\,467\,$\times$$10^{-3}$ \\
13 & 1.090\,093\,959\,979 &   -1.471\,024\,$\times$$10^{-3}$ \\
14 & 1.090\,094\,186\,392 &    9.771\,970\,$\times$$10^{-4}$ \\
15 & 1.090\,094\,280\,906 &   -6.491\,561\,$\times$$10^{-4}$  \\
16 & 1.090\,094\,320\,376 &    4.312\,391\,$\times$$10^{-4}$ \\
17 & 1.090\,094\,336\,865 &   -2.864\,762\,$\times$$10^{-4}$  \\
18 & 1.090\,094\,343\,755 &    1.903\,092\,$\times$$10^{-4}$ \\
19 & 1.090\,094\,346\,634 &   -1.264\,245\,$\times$$10^{-4}$ \\
20 & 1.090\,094\,347\,837 &    8.398\,518\,$\times$$10^{-5}$ \\
21 & 1.090\,094\,348\,340 &   -5.579\,230\,$\times$$10^{-5}$
\end{tabular}
\end{table}
\end{minipage}

We now study the asymptotic scaling behavior of the
period-doubling sequences in both subsystems near the bicritical
point. The scaling behavior in the first subsystem is obviously
the same as that in the 1D maps \cite{Feigenbaum}. That is, the
sequences $\{ A_n \}$ and $\{ x_n \}$ accumulate to their limit
values, $A=A_c$ $(=1.401\,155\,189\,092 \cdots)$ and $x=0$,
geometrically as follows:
\begin{equation}
A_n - A_c \sim \delta_1^{-n},\;\; x_n  \sim \alpha_1^{-n}\;\;
{\rm for\;large\;} n.
\label{eq:SF1}
\end{equation}
The scaling factors $\delta_1$ and $\alpha_1$ are just the
Feigenbaum constants $\delta$ $(=4.669 \cdots)$
and $\alpha$ $(=-2.502 \cdots)$ for the 1D maps, respectively.
However, the second subsystem exhibits a non-Feigenbaum critical
behavior, unlike the case of the first subsystem. The two sequences
$\{ B_n \}$ and $\{ y_n \}$ also converge geometrically to their limit
values $B=B_c$ $(=1.090\,094\,348\,701)$ and $y=0$, respectively,
where the value of $B_c$ is obtained using the superconverging method
\cite{Mackay}. To obtain the convergence rates of the two sequences,
we define the scaling factors of level $n$:
\begin{equation}
\delta_{2,n} \equiv { {B_{n-1}-B_n} \over {B_n-B_{n+1}} },\;\;
\alpha_{2,n} \equiv { {y_{n-1}-y_n} \over {y_n-y_{n+1}} }.
\label{eq:SF2n}
\end{equation}
These two sequences $\{ \delta_{2,n} \}$ and $\{ \alpha_{2,n} \}$ are
listed in Table \ref{tab:SF2}, and they converge to their limit values,
\begin{equation}
\delta_2 \simeq 2.3928, \;\;  \alpha_2 \simeq -1.5053,
\label{eq:SF2}
\end{equation}
respectively. Note that these scaling factors are completely different
from those in the first subsystem (i.e., the Feigenbaum constants
for the 1D maps).

\noindent
\begin{minipage}{\columnwidth}
\begin{table}
\caption{Sequences of the parameter and orbital scaling factors,
         $\{ \delta_{2,n} \}$ and $\{ \alpha_{2,n} \}$, in the
         second subsystem.
         }
\label{tab:SF2}
\begin{tabular}{ccc}
    $n$ & $\delta_{2,n}$ & $\alpha_{2,n}$   \\
\tableline
10 & 2.429\,8 & -1.505\,733\,1 \\
11 & 2.413\,7 & -1.505\,515\,4 \\
12 & 2.406\,3 & -1.505\,428\,1 \\
13 & 2.398\,8 & -1.505\,375\,3 \\
14 & 2.395\,6 & -1.505\,344\,0 \\
15 & 2.394\,6 & -1.505\,331\,6 \\
16 & 2.393\,7 & -1.505\,325\,6 \\
17 & 2.393\,1 & -1.505\,321\,5 \\
18 & 2.393\,0 & -1.505\,319\,8 \\
19 & 2.392\,9 & -1.505\,319\,1 \\
20 & 2.392\,8 & -1.505\,318\,6
\end{tabular}
\end{table}
\end{minipage}

For evidence of scaling, we compare the chaotic attractors, shown
in Fig.~\ref{fig:CA}, for the three values of $(A,B)$ near the
bicritical point $(A_c,B_c)$. All these attractors are the
hyperchaotic ones with two positive Lyapunov exponents \cite{HC},
\begin{eqnarray}
\sigma_1 &=& \lim_{m \rightarrow \infty} {1 \over m} \sum_{t=1}^{m}
  \ln |2A x_t|,\;\;
\sigma_2 = \lim_{m \rightarrow \infty} {1 \over m} \sum_{t=1}^{m}
   \ln |2B y_t|. \nonumber \\
   \label{eq:Lexp}
\end{eqnarray}
Here the first and second Lyapunov exponents $\sigma_1$ and
$\sigma_2$ denote the average exponential divergence rates of
nearby orbits in the the first and second subsystems,
respectively. Figure \ref{fig:CA}(a) shows the hyperchaotic
attractor with $\sigma_1 \simeq 0.242$ and $\sigma_2 \simeq 0.04$
for $A=A_c + \Delta A$ and $B=B_c + \Delta B$, where $\Delta A =
\Delta B =0.1$. This attractor consists of two pieces. To see
scaling, we first rescale $\Delta A$ and $\Delta B$ with the
parameter scaling factors $\delta_1$ and $\delta_2$, respectively.
The attractor for the rescaled parameter values of $A=A_c + \Delta
A / \delta_1$ and $B=B_c + \Delta B / \delta_2$ is shown in
Fig.~\ref{fig:CA}(b). It is also the hyperchaotic attractor with
$\sigma_1 \simeq 0.121$ and $\sigma_2 \simeq 0.02$. We next
magnify the region in the small box (containing the origin) by the
scaling factor $\alpha_1$ for the $x$ axis and $\alpha_2$ for the
$y$ axis, and then we get the picture in Fig.~\ref{fig:CA}(c).
Note that the picture in Fig.~\ref{fig:CA}(c) reproduces the
previous one in Fig.~\ref{fig:CA}(a) approximately. Repeating the
above procedure once more, we obtain the two pictures in
Figs.~\ref{fig:CA}(d) and \ref{fig:CA}(e). That is,
Fig.~\ref{fig:CA}(d) shows the hyperchaotic attractor with $\sigma
\simeq 0.061$ and $\sigma_2 \simeq 0.01$ for $A=A_c + \Delta A /
\delta_1^2$ and $B=B_c + \Delta B / \delta_2^2$. Magnifying the
region in the small box with the scaling factors $\alpha_1^2$ for
the $x$-axis and $\alpha_2^2$ for the $y$-axis, we also obtain the
picture in Fig.~\ref{fig:CA}(e), which reproduces the previous one
in Fig.~\ref{fig:CA}(c) with an increased accuracy.

\noindent
\begin{minipage}{\columnwidth}
\begin{figure}
\vspace*{-0.5cm}
\centerline{ 
 \epsfig{file={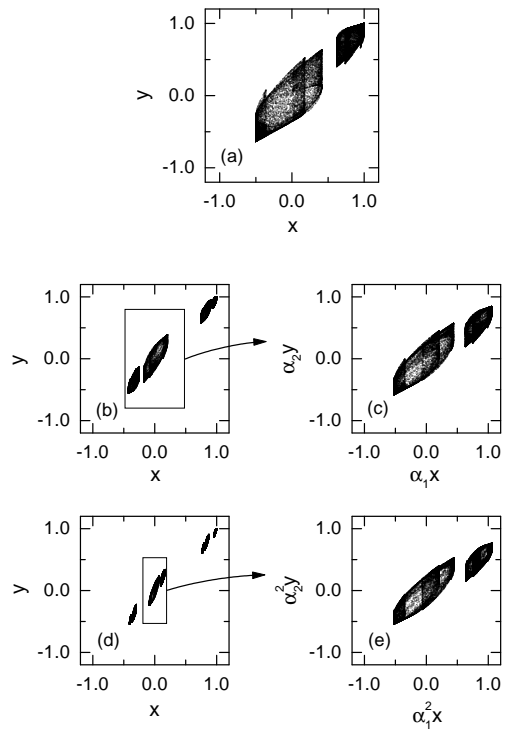}, width=\columnwidth}}
\vspace{-1.5cm}
\caption{Hyperchaotic attractors for the three values of $(A,B)$ near
  the bicritical point $(A_c.B_c)$; in (a) $(A,B)=(A_c+\Delta A, B_c+
  \Delta B)$ $(\Delta A = \Delta B =0.1)$, in (b) and (c) $(A,B) =
  (A_c+\Delta A/ \delta_1, B_c+\Delta B/ \delta_2)$, and in (d) and
  (e) $(A,B)=(A_c+ \Delta A / \delta_1^2, B_c+\Delta B/ \delta_2^2)$.
  The picture in (c) is obtained by magnifying the region in the small
  box in (b) with the scaling factors $\alpha_1$ for the $x$-axis and
  $\alpha_2$ for the $y$-axis. Similarly, we also obtain the
  picutre (e) by magnifying the region inside the small box in (d)
  with the scaling factors $\alpha_1^2$ for the $x$-axis and
  $\alpha_2^2$ for the $y$-axis. Comparing the pictures in (a), (c),
  and (e), one can see that each successive magnified picture
  reproduces the previous one with an accuracy with the depth of
  resolution.
     }
\label{fig:CA}
\vspace{0.5cm}
\end{figure}
\end{minipage}

So far we have seen the scaling near the bicritical point, and now
turn to a discussion of the behavior exactly at the bicritical point
$(A_c,B_c)$. There exist an infinity of unstable periodic orbits with
period $2^n$ at the bicritical point. The orbit points $x_n$ and
$y_n$, approaching the zero in the first and second subsystems, vary
asymptotically in proportion to $\alpha_1^{-n}$ and $\alpha_2^{-n}$,
respectively. The stability multipliers $\lambda_{1,n}$ and
$\lambda_{2,n}$ of the orbits with period $2^n$ also converge to the
critical stability multipliers $\lambda_1^*$ and $\lambda_2^*$,
respectively. Here $\lambda_1^*$ $(=-1.601\,191\,\cdots)$ in the first
subsystem is just the critical stability multiplier for the case of
the 1D maps \cite{Feigenbaum}. However, as listed in Table
\ref{tab:SM2}, the second subsystem has the different critical
stability multiplier,
\begin{equation}
\lambda_2^* = -1.178\,85\,\cdots.
\end{equation}
Consequently, the periodic orbits at the bicritical point have the
same stability multipliers $\lambda_1^*$ and $\lambda_2^*$ for
sufficiently large $n$.

\noindent
\begin{minipage}{\columnwidth}
\begin{table}
\caption{Sequences of the second stability multipliers,
         $\{ \lambda_{2,n} \}$ of the orbits with period $2^n$ at the
         bicritical point.
         }
\label{tab:SM2}
\begin{tabular}{cc}
    $n$ & $\lambda_{2,n}$   \\
\tableline
10 & -1.178\,829 \\
11 & -1.178\,842 \\
12 & -1.178\,839 \\
13 & -1.178\,850 \\
14 & -1.178\,855 \\
15 & -1.178\,854 \\
16 & -1.178\,854 \\
17 & -1.178\,855 \\
18 & -1.178\,855 \\
19 & -1.178\,854
\end{tabular}
\end{table}
\end{minipage}

\section{Renormalization Group Analysis of The Bicritical Behavior}
\label{sec:RA}
Employing two different methods, we make the RG analysis of the
bicritical behavior. We first use the truncation method, and
analytically obtain the corresponding fixed point and its
relevant eigenvalues. These analytic results are not bad as the
lowest-order approximation. To improve the accuracy, we also use the
numerical eigenvalue-matching method, and obtain the bicritical point,
the parameter and orbital scaling factors, and the critical stability
multipliers. Note that the accuracy in the numerical RG results is
improved remarkably with increasing the level $n$.

\subsection{Truncation Method}
In this subsection, employing the truncation method \cite{TM}, we
analytically make the RG analysis of the bicritical behavior in the
unidirectionally-coupled map T of the form,
\begin{equation}
T:\; x_{t+1}= f(x_t),\;\;y_{t+1}=g(x_t,y_t),
\label{eq:UCM1}
\end{equation}
where $x_t$ and $y_t$ are the state variables at a discrete time $t$
in the first and second subsystems, respectively.
Truncating the map (\ref{eq:UCM1}) at its quadratic terms, we have
\begin{equation}
T_{{\bf P}}:\;x_{t+1}={a \over b} + b x_t^2,\;\;
           y_{t+1}={ c \over d} + d y_t^2 + {e \over d} x_t^2,
\label{eq:UCM2}
\end{equation}
which is a five-parameter family of unidirecionally-coupled maps.
$\bf P$ represents the five parameters, i.e., ${\bf P} = (a,b,c,d,e)$.
The construction of Eq.~(\ref{eq:UCM2}) corresponds to a truncation
of the infinite dimensional space of unidirectionally-coupled maps
to a five-dimensional space. The parameters $a, b, c, d, $ and $e$ can
be regarded as the coordinates of the truncated space. We also note
that this truncation method corresponds to the lowest-order polynomial
approximation.

We look for fixed points of the renormalization operator ${\cal R}$ in
the truncated five-dimensional space of unidirectionally-coupled maps,
\begin{equation}
{\cal R} (T) = \Lambda T^2 \Lambda^{-1}.
\label{eq:RO}
\end{equation}
Here the rescaling operator $\Lambda$ is given by
\begin{equation}
\Lambda = \left (
\begin{array}{cc}
\alpha_1 & 0 \\
0 & \alpha_2
\end{array}
\right ),
\label{eq:SO}
\end{equation}
where $\alpha_1$ and $\alpha_2$ are the rescaling factors in the first
and second subsystems, respectively.

The operation ${\cal R}$ in the truncated space can be represented by
a transformation of parameters, i.e., a map from ${\bf P} \equiv
(a,b,c,d,e)$ to ${\bf P^{\prime}} \equiv
(a^{\prime},b^{\prime},c^{\prime},d^{\prime},e^{\prime}),$
\begin{mathletters}
\label{eq:PT}
\begin{eqnarray}
a^{\prime}&=& 2 a^2 (1+a),  \label{eq:RGTA} \\
b^{\prime}&=& {2 \over \alpha_1} a b,  \label{eq:RGTB} \\
c^{\prime}&=& 2c (c+c^2+ e {a^2 \over b^2}),  \label{eq:RGTC} \\
d^{\prime}&=& {2 \over \alpha_2} c d,  \label{eq:RGTD} \\
e^{\prime}&=& {4 \over \alpha_1^2} c e (a+c).
\label{eq:RGTE}
\end{eqnarray}
\end{mathletters}
The fixed point ${\bf P}^* =(a^*,b^*,c^*,d^*,e^*)$ of this map can be
determined by solving ${\bf P}^{\prime}={\bf P}$. The parameters $b$
and $d$ set only the scales in the $x$ and $y$, respectively, and thus
they are arbitrary. We now fix the scales in $x$ and $y$ by setting
$b=d=1$. Then, we have, from Eqs.(\ref{eq:RGTA})-(\ref{eq:RGTE}), five
equations for the five unknowns $\alpha_1, a^*, \alpha_2, c^*,$ and
$e^*$. We thus find one solution, associated with the bicritical
behavior, as will be seen below. The map (\ref{eq:UCM2}) with a
solution ${\bf P^*}$ $(T_{{\bf P^*}})$ is the fixed map of the
renormalization transformation ${\cal R}$; for brevity $T_{{\bf P^*}}$
will be denoted as $T^*$.

We first note that Eqs.~(\ref{eq:RGTA})-(\ref{eq:RGTB}) are only for
the unknowns $\alpha_1$ and $a^*$. We find one solution for $\alpha_1$
and $a^*$, associated with the period-doubling bifurcation in the
first subsystem,
\begin{equation}
 \alpha_1=-1-\sqrt{3}=-2.732 \cdots,\;a^*={\alpha_1 \over 2}.
 \label{eq:alpha1}
\end{equation}
Substituting the values for $\alpha_{1}$ and $a^*$ into
Eqs.~(\ref{eq:RGTC})-(\ref{eq:RGTE}), we obtain one solution for
$\alpha_2$, $c^*$, and $e^*$, associated with the bicriticality,
\begin{mathletters}
\begin{eqnarray}
\alpha_2 &=& {1 \over 2} (1 + \sqrt{3} - \sqrt{5} - \sqrt{15})= -1.688
\cdots,\label{eq:alpha2} \\
c^*&=&{\alpha_2 \over 2},\;\;
e^*=1+{1 \over 2}(\sqrt{15}-3 \sqrt{3})=0.338 \cdots.
\end{eqnarray}
\end{mathletters}
Compared with the values, $\alpha_1=-2.502\cdots$ and $\alpha_2=-1.505
\cdots$, obtained by a direct numerical method, the analyitc results
for $\alpha_1$ and $\alpha_2$, given in Eqs.~(\ref{eq:alpha1}) and
(\ref{eq:alpha2}), are not bad as the lowest-order approximation.

Consider an infinitesimal perturbation $\epsilon \, \delta {\bf P}$ to
a fixed point ${\bf P}^*$ of the transformation of parameters
(\ref{eq:RGTA})-(\ref{eq:RGTE}). Linearizing the transformation at
${\bf P}^*$, we obtain the equation for the evolution of $\delta
{\bf P}$,
\begin{equation}
\delta {\bf P}^{\prime}= J \delta {\bf P},
\end{equation}
where $J$ is the Jacobian matrix of the transformation at ${\bf P}^*$.

The $5 \times 5$ Jacobian matrix $J$ has a semi-block form, because
we are considering the unidirectionally-coupled case. Therefore,
one can easily obtain its eigenvalues. The first two eigenvalues,
associated with the first subsystem, are those of the following
$2 \times 2$ matrix,
\begin{eqnarray}
\left. M_1 = {\frac{ {\partial (a^{\prime},b^{\prime})} }{{\partial
(a,b)} }} \right|_{{\bf P^*}} = \left(
\begin{array}{cc}
3-\alpha_1 & 0 \\
{2 / \alpha_1} & 1
\end{array}
\right).
\label{eq:M_1}
\end{eqnarray}
Hence the two eigenvalues of $M_1$, $\delta_1$ and $\delta_1'$, are
given by
\begin{equation}
 \delta_1 = 4+\sqrt{3}=5.732 \cdots,\;\;\delta_1'=1.
 \label{eq:delta1}
\end{equation}
Here the relevant eigenvalue $\delta_1$ is associated with the
scaling of the control parameter in the first subsystem, while
the marginal eigenvalue $\delta_1'$ is associated with the
scale change in $x$. When compared with the numerical value,
$\delta_1$ $(=4.669 \cdots)$, the analytic result for $\delta_1$,
given in Eq.~(\ref{eq:delta1}), is not bad as the lowest-order
approximation.

The remaining three eigenvalues, associated with the second subsystem,
are those of the following $3 \times 3$ matrix,

\end{multicols}

\vspace{0.3cm}
\begin{equation}
\left. M_2 = {\frac{ {\partial (c^{\prime},d^{\prime},e^{\prime})} }
{{\partial (c,d,e)} }} \right|_{{\bf P^*}}
= \left(
\begin{array}{ccc}
4c^* + 6c^{*2}+2a^{*2} e^* & 0 & 2 a^{*2} c^*\\
2/ \alpha_2 & 2c^* / \alpha_2 & 0 \\
4e^* (a^*+2c^*) / \alpha_1^2 & 0 & 4c^* (a^*+c^*) / \alpha_1^2
\end{array}
\right).
\label{eq:M_2}
\end{equation}

\begin{multicols}{2}
The three eigenvalues of $M_2$, $\delta_2$, $\delta_2'$, and
$\delta_2''$, are given by
\begin{mathletters}
\begin{eqnarray}
\delta_2 &=& (u+\sqrt{v})/2=3.0246\cdots, \label{eq:delta2} \\
\delta_2' &=& (u-\sqrt{v})/2=0.1379 \cdots,\;\;
\delta_2''=1,
\end{eqnarray}
\end{mathletters}
where
\begin{mathletters}
\begin{eqnarray}
u&=&(17+7 \sqrt{3} - 5 \sqrt{5} -3 \sqrt{15})/2, \\
v&=&104+53\sqrt{3} - 44 \sqrt{5} -23 \sqrt{15}.
\end{eqnarray}
\label{eq:uv}
\end{mathletters}
The first eigenvalue $\delta_2$ is a relevant eigenvalue, associated
with the scaling of the control parameter in the second subsystem, the
second eigenvalue $\delta_2'$ is an irrelavant one, and the third
eigenvalue $\delta_2''$ is a marginal eigenvalue, associated with the
scale change in $y$. We also compare the analytic result for
$\delta_2$, given in Eq.~(\ref{eq:delta2}), with the numerical value
$(\delta_2 \simeq 2.3928)$ in Eq.~(\ref{eq:SF2}), and find that the
analytic one is not bad as the lowest-order approximation.

As shown in Sec.~\ref{sec:SB}, stability multipliers of an orbit
with period $2^n$ at the bicritical point converge to the critical
stability multipliers, $\lambda_1^*$ $(=-1.601 \cdots)$ and
$\lambda_2^*$ $(=-1.178 \cdots)$ as $n \rightarrow \infty$. We now
obtain these critical stability analytically. The invariance of
the fixed map $T^*$ under the renormalization transformation $\cal
R$ implies that, if $T^*$ has a periodic point $(x,y)$ with period
$2^n$, then $\Lambda^{-1} (x,y)$ is a periodic point of $T^*$ with
period $2^{n+1}$. Since rescaling does not affect the stability
multipliers, all the orbits with period $2^n$ $(n=0,1,2,\dots)$
have the same stability multipliers, which are just the critical
stability multipliers $\lambda_1^*$ and $\lambda_2^*$. That is,
the critical stability multipliers have the values of the
stability multipliers of the fixed point $(x^*,y^*)$ of the fixed
map $T^*$,
\begin{equation}
\lambda_1^* = 2 x^* = -1.5424 \cdots,\;\; \lambda_2^*
= 2 y^* = -0.8899\cdots,
\label{eq:CSM}
\end{equation}
where
\begin{mathletters}
\begin{eqnarray}
x^* &=& (1 - \sqrt{3 + 2 \sqrt{3}})/2,\;\;
y^*= (1- \sqrt{w})/2, \\
w &=& 5+ 3 \sqrt{3} -2 \sqrt{5} - \sqrt{15} \nonumber \\
&& + \sqrt{3 + 2 \sqrt{3}}\; (2 - 3 \sqrt{3} + \sqrt{15}).
\end{eqnarray}
\label{eq:FP1}
\end{mathletters}
We also note that the analytic values for $\lambda^{*}_{1}$ and
$\lambda^{*}_{2}$ are not bad, when compared with their numerical
values.

\subsection{Eigenvalue-Matching Method}
In this subsection, we employ the eigenvalue-matching method \cite{EM}
and numerically make the RG analysis of the bicritical behavior in the
unidirectionally-coupled map $T$ of Eq.~(\ref{eq:UCM}). As the level
$n$ increases, the accuracy in the numerical RG results are remarkably
improved.

The basic idea is to associate a value $(A',B')$ for each value
$(A,B)$ such that $T^{(n+1)}_{(A',B')}$ locally resembles
$T^{(n)}_{(A,B)}$, where $T^{(n)}$ is the $2^n$th-iterated map of
$T$ (i.e., $T^{(n)}=T^{2^n}$). A simple way to implement this idea
is to linearize the maps in the neighborhood of their respective
fixed points and equate the corresponding eigenvalues.

Let $\{ z_t \}$ and $\{ z'_t \}$ be two successive cycles of period
$2^n$ and $2^{n+1}$, respectively, i.e.,
\begin{equation}
z_t = T^{(n)}_{(A,B)} (z_t),\;\; z'_t = T^{(n+1)}_{(A',B')} (z'_t);
\;z_t = (x_t,y_t).
\label{eq:FP2}
\end{equation}
Here $x_t$ depends only on $A$, but $y_t$ is dependent on both
$A$ and $B$, i.e., $x_t= x_t(A)$ and $y_t=y_t(A,B)$.
Then their linearized maps at $z_t$ and $z'_t$ are given by
\begin{mathletters}
\begin{eqnarray}
DT^{(n)}_{(A,B)} &=& \prod_{t=1}^{2^n} DT_{(A,B)}(z_t), \\
DT^{(n+1)}_{(A',B')} &=& \prod_{t=1}^{2^{n+1}} DT_{(A',B')}(z'_t).
\end{eqnarray}
\label{eq:LM}
\end{mathletters}
(Here $DT$ is the linearized map of $T$.)
Let their eigenvalues, called the stability multipliers, be
($\lambda_{1,n}(A), \lambda_{2,n}(A,B))$ and
($\lambda_{1,n+1}(A'), \lambda_{2,n+1}(A',B'))$.
The recurrence relations for the old and new parameters are then
given by equating the stability multipliers of level $n$,
$\lambda_{1,n}(A)$ and $\lambda_{2,n}(A,B)$, to those of the next
level $n+1$, $\lambda_{1,n+1}(A')$ and $\lambda_{2,n+1}(A',B')$, i.e.,
\begin{mathletters}
\begin{eqnarray}
\lambda_{1,n}(A) &=& \lambda_{1,n+1}(A'), \\
\lambda_{2,n}(A,B) &=& \lambda_{2,n+1}(A',B').
\end{eqnarray}
\label{eq:PRR}
\end{mathletters}

The fixed point $(A^*,B^*)$ of the renormalization transformation
(\ref{eq:PRR}),
\begin{mathletters}
\begin{eqnarray}
\lambda_{1,n}(A^*) &=& \lambda_{1,n+1}(A^*), \\
\lambda_{2,n}(A^*,B^*) &=& \lambda_{2,n+1}(A^*,B^*).
\end{eqnarray}
\label{eq:BCP}
\end{mathletters}
gives the bicritical point $(A_c,B_c)$. By linearizing the
renormalization transformation (\ref{eq:PRR}) at the fixed point
$(A^*,B^*)$, we have
\begin{eqnarray}
\left( \begin{array}{c}
                    \Delta A \\
                    \Delta B
                   \end{array}
           \right)
&=&  \left(
\begin{array}{cc}
\left. {\partial A \over \partial A'} \right|_{*} &
\left. {\partial A \over \partial B'} \right|_{*} \\
\left. {\partial B \over \partial A'} \right|_{*} &
\left. {\partial B \over \partial B'} \right|_{*}
\end{array}
\right )
\left( \begin{array}{c}
                    \Delta A' \\
                    \Delta B'
                   \end{array}
           \right) \\
&=& \Delta_n
\left( \begin{array}{c}
                    \Delta A' \\
                    \Delta B'
                   \end{array}
\right),
\end{eqnarray}
where $\Delta A= A-A^*$, $\Delta B= B-B^*$, $\Delta A'= A'-A^*$,
$\Delta B'= B'-B^*$, and
\begin{mathletters}
\begin{eqnarray}
\Delta_n &=& \Gamma_n^{-1} \Gamma_{n+1};
\label{eq:Delta} \\
\Gamma_n &=& \left (
\begin{array}{cc}
\left. {{d\lambda_{1,n} \over dA}} \right|_{*} & 0 \\
\left. {{\partial \lambda_{2,n} \over \partial A}} \right|_{*} &
\left. {{\partial \lambda_{2,n} \over \partial B}} \right|_{*}
\label{eq:Gamma}
\end{array}
\right ), \\
\Gamma_{n+1} &=& \left (
\begin{array}{cc}
\left. {{d\lambda_{1,n+1} \over dA'}} \right|_{*} & 0 \\
\left. {{\partial \lambda_{2,n+1} \over \partial A'}} \right|_{*}  &
\left. {{\partial \lambda_{2,n+1} \over \partial B'}} \right|_{*}
\end{array}
\right ).
\end{eqnarray}
\end{mathletters}
Here $\Gamma_n^{-1}$ is the inverse of $\Gamma_n$ and the asterisk
denotes the fixed point $(A^*,B^*)$. After some algebra, we obtain
the analytic formulas for the eigenvalues $\delta_{1,n}$ and
$\delta_{2,n}$ of the matrix $\Delta_n$,
\begin{mathletters}
\begin{eqnarray}
\delta_{1,n} &=&
 { {\left . {d \lambda_{1,n+1} \over d A'} \right|_*} \over
  {\left . {d \lambda_{1,n} \over d A} \right|_*} },
  \label{eq:EV1} \\
  \delta_{2,n} &=&
 { {\left . {\partial \lambda_{2,n+1} \over \partial B'} \right|_*}
 \over   {\left . {\partial \lambda_{2,n} \over \partial B}
  \right|_*} }.
\label{eq:EV2}
\end{eqnarray}
\label{eq:PSF}
\end{mathletters}
As $n \rightarrow \infty$ $\delta_{1,n}$ and $\delta_{2,n}$ approach
$\delta_1$ and $\delta_2$, which are just the parameter scaling
factors in the first and second subsystems, respectively.
Note also that as in the 1D case, the local rescaling factors of the
state variables are simply given by
\begin{mathletters}
\begin{eqnarray}
\alpha_{1,n} &=&  {\left. {dx \over dx'} \right|_*}
              = { \delta_{1,n} \over t_{1,n} },
\label{eq:LSF1}  \\
\alpha_{2,n} &=&  {\left. {dy \over dy'} \right|_*}
              = { \delta_{2,n} \over t_{2,n} },
\label{eq:LSF2}
\end{eqnarray}
\label{eq:OSF}
\end{mathletters}
where
\begin{equation}
t_{1,n} =
 { {\left . {d x' \over d A'} \right|_*} \over
  {\left . {d x \over d A} \right|_*} }, \;
  t_{2,n} =
 { {\left . {\partial y' \over \partial B'} \right|_*}
 \over   {\left . {\partial y \over \partial B}
  \right|_*}}.
\end{equation}
Here $\alpha_{1,n}$ and $\alpha_{2,n}$ also converge to the orbital
scaling factors, $\alpha_1$ and $\alpha_2$, in the first and second
subsystems, respectively.

We numerically follow the orbits with period $2^n$ in the
unidirectionally-coupled maps $T$ of Eq.~(\ref{eq:UCM}) and make
the RG analysis of the bicritical behavior. Some results for the
intermediate level $n$ are shown in Fig.~\ref{fig:RA}. Figure
\ref{fig:RA}(a) shows the plots of the first stability multiplier
$\lambda_{1,n}(A)$ versus $A$ for the cases $n=6, 7$. We note that
the intersection point, denoted by the solid circle, of the two
curves $\lambda_{1,6}$ and $\lambda_{1,7}$ gives the point
$(A^*_6, \lambda_{1,6}^*)$ of level $6$, where $A^*_6$ and
$\lambda^*_{1,6}$ are the critical point and the critical
stability multiplier, respectively, in the first subsystem. As the
level $n$ increases, $A^*_n$ and $\lambda^*_{1,n}$ approach their
limit values $A^*$ and $\lambda_1^*$, respectively. Note also that
the ratio of the slopes of the curves, $\lambda_{1,6}(A)$ and
$\lambda_{1,7}(A)$, for $A=A^*_6$ gives the parameter scaling
factor $\delta_{1,6}$ of level $6$ in the first subsystem.
Similarly, Fig.~\ref{fig:RA}(b) shows the plots of the second
stability multiplier $\lambda_{2,n}(A^*_6,B)$ versus $B$ for the
cases $n=6, 7$. The intersection point, denoted also by the solid
circle, of the two curves $\lambda_{2,6}(A^*_6,B)$ and
$\lambda_{2,7}(A^*_6,B)$ gives the point $(B^*_6,\lambda_{2,6}^*)$
of level $6$, where $B^*_6$ and $\lambda_{2,6}^*$ are the critical
point and the critical stability multiplier, respectively, in the
second subsystem. As the level $n$ increases, $B^*_n$ and
$\lambda^*_{2,n}$ also converge to their limit values, $B^*$ and
$\lambda_2^*$, respectively. As in the first subsystem, the ratio
of the slopes of the curves, $\lambda_{2,6}(A^*_6,B)$ and
$\lambda_{2,7}(A^*_6,B)$, for $B=B^*_6$ gives the parameter
scaling factor $\delta_{2,6}$ of level $6$ in the second
subsystem.

\noindent
\begin{minipage}{\columnwidth}
\begin{figure}
\centerline{ 
 \epsfig{file={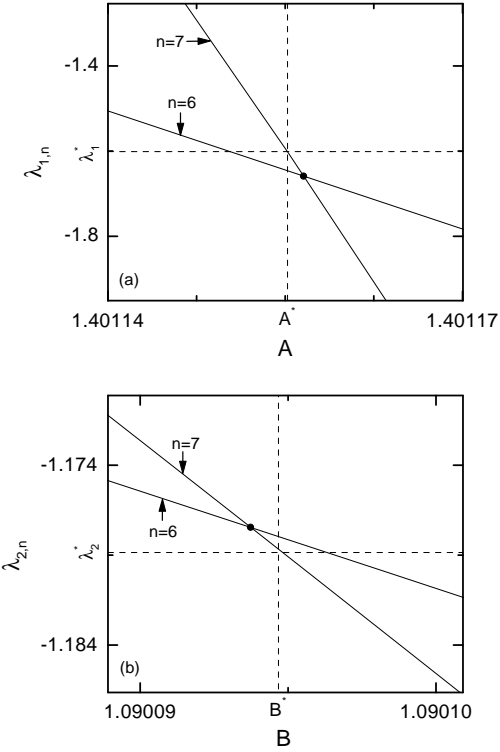}, width=\columnwidth}}
\vspace{-1cm}
\caption{Plots of (a) the first stability multipliers
$\lambda_{1,n}(A)$ versus $A$ and (b) the second stability
multipliers $\lambda_{2,n}(A^*_6,B)$ versus $B$ for the cases
$n=6,7$. In (a), the intersection point, denoted by the solid
circle, of the two curves $\lambda_{1,6}$ and $\lambda_{1,7}$
gives the point $(A^*_6,\lambda_{1,6}^*)$ of level $6$. As $n
\rightarrow \infty$, $(A^*_n,\lambda^*_{1,n})$ converges to its
limit point $(A^*,\lambda_1^*)$. Similarly, in (b), the
intersection point, denoted also by the solid circle, of the two
successive curves $\lambda_{2,6}(A^*_6,B)$ and
$\lambda_{2,7}(A^*_6,B)$ gives the point $(B^*_6,\lambda_{2,6}^*)$
of level $6$. As $n \rightarrow \infty$, $(B^*_n,\lambda^*_{2,n})$
also approaches its limit point $(B^*,\lambda_2^*)$. For other
details, see the text.
     }
\label{fig:RA}
\end{figure}
\end{minipage}

With increasing the level up to $n=15$, we numerically make the RG
analysis of the bicritical behavior. We first solve
Eq.~(\ref{eq:BCP}) and obtain the bicritical point $(A^*_n,B^*_n)$
of level $n$ and the pair of critical stability multipliers
$(\lambda^*_{1,n}, \lambda^*_{2,n})$ of level $n$. Next, we use
the formulas of Eqs.~(\ref{eq:PSF}) and (\ref{eq:OSF}) and obtain
the parameter and orbital scaling factors of level $n$,
respectively. These numerical RG results for the first and second
subsystems are listed in Tables \ref{tab:RAA} and \ref{tab:RAB},
respectively. Note that the accuracy in the numerical RG results
is remarkably improved with the level $n$ and their limit values
agree well with those obtained by a direct numerical method.
\end{multicols}

\begin{center}
\widetext
\noindent
\begin{minipage}{14cm}
\begin{table}
\caption{Sequences of the critical point, the first critical
         stability multiplier, the parameter and orbital scaling
         factors, $\{ A^*_n \}$, $\{ \lambda^*_{1,n} \}$,
         $\{ \delta_{1,n} \}$ and $\{ \alpha_{1,n} \}$, in the
         first subsystem. For comparison, we also list the results
         obtained by a direct numerical method in the last row.
         }
\label{tab:RAA}
\begin{tabular}{ccccc}
    $n$ & $A^*_n$ & $\lambda^*_{1,n}$ & $\delta_{1,n}$ & $\alpha_{1,n}$ \\
\tableline
6 & 1.401\,155\,189\,088\,929\,1 & -1.601\,191\,211\,121\,2 &
   4.669\,203\,072\,1 & -2.502\,620\,459\,5 \\
7 & 1.401\,155\,189\,092\,133\,2 & -1.601\,191\,342\,517\,1 &
   4.669\,201\,428\,5 & -2.502\,845\,988\,3 \\
8 & 1.401\,155\,189\,092\,048\,4 & -1.601\,191\,326\,288\,7 &
   4.669\,201\,631\,4 & -2.502\,894\,652\,0 \\
9 & 1.401\,155\,189\,092\,050\,7 & -1.601\,191\,328\,294\,3 &
   4.669\,201\,606\,3 & -2.502\,905\,037\,7 \\
10 & 1.401\,155\,189\,092\,050\,6 & -1.601\,191\,328\,046\,4 &
   4.669\,201\,609\,4 & -2.502\,907\,267\,8 \\
11 & 1.401\,155\,189\,092\,050\,6 & -1.601\,191\,328\,077\,0 &
   4.669\,201\,609\,1 & -2.502\,907\,744\,9 \\
12 & 1.401\,155\,189\,092\,050\,6 & -1.601\,191\,328\,073\,2 &
   4.669\,201\,609\,1 & -2.502\,907\,847\,2 \\
13 & 1.401\,155\,189\,092\,050\,6 & -1.601\,191\,328\,073\,7 &
   4.669\,201\,609\,1 & -2.502\,907\,869\,1 \\
14 & 1.401\,155\,189\,092\,050\,6 & -1.601\,191\,328\,073\,6 &
   4.669\,201\,609\,1 & -2.502\,907\,873\,8 \\
15 & 1.401\,155\,189\,092\,050\,6 & -1.601\,191\,328\,073\,6 &
   4.669\,201\,609\,1 & -2.502\,907\,874\,8 \\
 & 1.401\,155\,189\,092\,050\,6 & -1.601\,191\,328\,073\,6 &
 4.669\,201\,609\,1 & -2.502\,907\,875\,1
\end{tabular}
\end{table}
\end{minipage}
\end{center}

\twocolumn
\narrowtext
\noindent
\begin{minipage}{\columnwidth}
\begin{table}
\caption{Sequences of the critical point, the second critical
         stability multiplier, the parameter and orbital scaling
         factors, $\{ B^*_n \}$, $\{ \lambda^*_{2,n} \}$,
         $\{ \delta_{2,n} \}$ and $\{ \alpha_{2,n} \}$, in the
         second subsystem. For comparison, we also list the results
         obtained by a direct numerical method in the last row.
         }
\label{tab:RAB}
\begin{tabular}{ccccc}
    $n$ & $B^*_n$ & $\lambda^*_{2,n}$ & $\delta_{2,n}$ & $\alpha_{2,n}$   \\
\tableline
6 & 1.090\,092\,490\,313 & -1.177\,467 & 2.395\,07 & -1.502\,785 \\
7 & 1.090\,094\,351\,702 & -1.178\,671 & 2.393\,58 & -1.503\,173 \\
8 & 1.090\,094\,321\,847 & -1.178\,625 & 2.393\,59 & -1.504\,426 \\
9 & 1.090\,094\,328\,376 & -1.178\,649 & 2.393\,10 & -1.504\,894 \\
10 & 1.090\,094\,347\,652 & -1.178\,820 & 2.392\,80 & -1.504\,993 \\
11 & 1.090\,094\,348\,817 & -1.178\,844 & 2.392\,81 & -1.505\,163 \\
12 & 1.090\,094\,348\,536 & -1.178\,830 & 2.392\,78 & -1.505\,263 \\
13 & 1.090\,094\,348\,675 & -1.178\,847 & 2.392\,74 & -1.505\,280 \\
14 & 1.090\,094\,348\,704 & -1.178\,856 & 2.392\,73 & -1.505\,296 \\
15 & 1.090\,094\,348\,701 & -1.178\,853 & 2.392\,73 & -1.505\,311 \\
   &  1.090\,094\,348\,701 & -1.178\,85 & 2.392\,7 &  -1.505\,318
\end{tabular}
\end{table}
\end{minipage}

\section{Summary}
\label{sec:Sum}
We have studied the scaling behavior of period doublings near the
bicritical point, corresponding to a threshold of chaos in both
subsystems. For this bicritical case, a new type of non-Feigenbaum
critical behavior appears in the second (response) subsystem, while
the first (drive) subsystem is in the Feigenbaum critical state.
Employing the truncation and eigenvalue-matching methods, we made the
RG analysis of the bicritical behavior. For the case of the truncation
method, we analytically obtained the fixed point, associated with the
bicritical behavior, and its relevant eigenvalues. These analytic RG
results are not bad as the lowest-order approximation. To improve the
accuracy, we also employed the numerical eigenvalue-matching RG
method, and obtained the bicritical point, the parameter and orbital
scaling factors, and the critical stability multipliers. Note that
the accuracy in the numerical RG results is improved remarkably with
increasing the level $n$. Consequently, these numerical RG results
agree well with the results obtained by a direct numerical method.
The results on the bicritical behavior in the abstract system of
unidirectionally-coupled 1D maps are also confirmed in the real
system of unidirectionally-coupled oscillators \cite{Kim}.

\acknowledgments
This work was supported by the Korea Research
Foundation under Project No. 1998-015-D00065 and by the Biomedlab
Inc. Some part of this manuscript was written during my visit to
the Center of Nonliner Studies in the Institute of Applied Physics
and Computational Mathematics, China, supported by the Korea
Science and Engineering Foundation and the National Natural
Science Foundation of China. I also thank Profs. Chen and Liu and
other members for their hospitality during my visit.

\end{document}